\documentclass[11pt,a4paper]{article}
\usepackage[T1]{fontenc}

\usepackage[a4paper,top=2cm,bottom=2cm,left=3cm,right=3cm,marginparwidth=1.75cm]{geometry}
 
\usepackage{amsmath}
\usepackage{graphicx}
\usepackage{hyperref}

\usepackage[affil-it]{authblk}

\makeatletter
\renewcommand{\section}{\@startsection {section}{1}{\z@}%
              {24pt}{12pt} {\large\scshape\bfseries}}
\renewcommand{\subsection}{\@startsection {subsection}{2}{\z@}%
             {12pt}{12pt}  {\itshape\bfseries}}

\usepackage{apacite}
\usepackage{natbib}

\usepackage{enumitem}

\newcommand{\keepcomment}{0} % 1 - Keep comments, 0 - Hide comments
 \usepackage[normalem]{ulem}
 \usepackage{xcolor}
\ifnum\keepcomment=1
	\usepackage[colorinlistoftodos,textsize=scriptsize]{todonotes} % todo
    \newcommand{\stkout}[1]{\ifmmode\text{\sout{\ensuremath{#1}}}\else\sout{#1}\fi}
    
\else
	
	\usepackage[disable]{todonotes} % todo
\fi

\newcommand{\todoi}[1]{\todo[inline]{#1}}
\usepackage{tabularx}
\usepackage{subcaption}
\usepackage{amsmath}

\usepackage{calrsfs} % For pazocal
\DeclareMathAlphabet{\pazocal}{OMS}{zplm}{m}{n}
\usepackage{amssymb}   % 提供 \mathbb
\usepackage{bm}        % 提供 \bm
\usepackage{algorithm}
\usepackage{algorithmic}
\usepackage{bbm} %For the indicator  function \mathbbm 1
\usepackage{hyperref}

\usepackage{multirow}

\usepackage{layout}

\usepackage{dblfloatfix}
\graphicspath{{../}}

\title{Real-Time Design of Public Transport Lines: Reconciling Adaptivity and Efficiency}

\author[1]{Duo Wang}
\author[2]{Andrea Araldo}
\author[3]{Mounim El-Yacoubi}

\affil[1]{PhD, Institut Polytechnique de Paris}
\affil[2]{Associate Professor, Institut Polytechnique de Paris}
\affil[3]{Professor, Institut Polytechnique de Paris}

\date{\vspace{-5ex}}

\begin{document}

\maketitle

\section*{Short Summary}

Demand-responsive transport (DRT) is typically routed by solving Dynamic Vehicle Routing Problems (DVRPs), where individual vehicle trajectories are adjusted on incoming requests. This limits demand consolidation and thus efficiency. On the other hand, Conventional Public Transport (CPT) bus systems are based on a network of lines and users find their routes on it, which provides high demand consolidation. However, such a network is built offline and cannot adapt to the demand.
We propose a public transport management strategy that reconciles efficiency and adaptivity by dynamically designing a structured network of lines via a receding-horizon optimization approach. Using real-world trip requests, we show that we nearly double the fraction of served requests compared to DVRP-based routing, and we serve more requests than CPT with lower user trip times. 
\, \\

\textbf{Keywords:} demand-responsive transport; public transport; network design; ride-sharing; dynamic systems

\section{Introduction}
\label{sec:introduction}

\begin{figure}
    \centering
    \includegraphics[width=.23\linewidth]{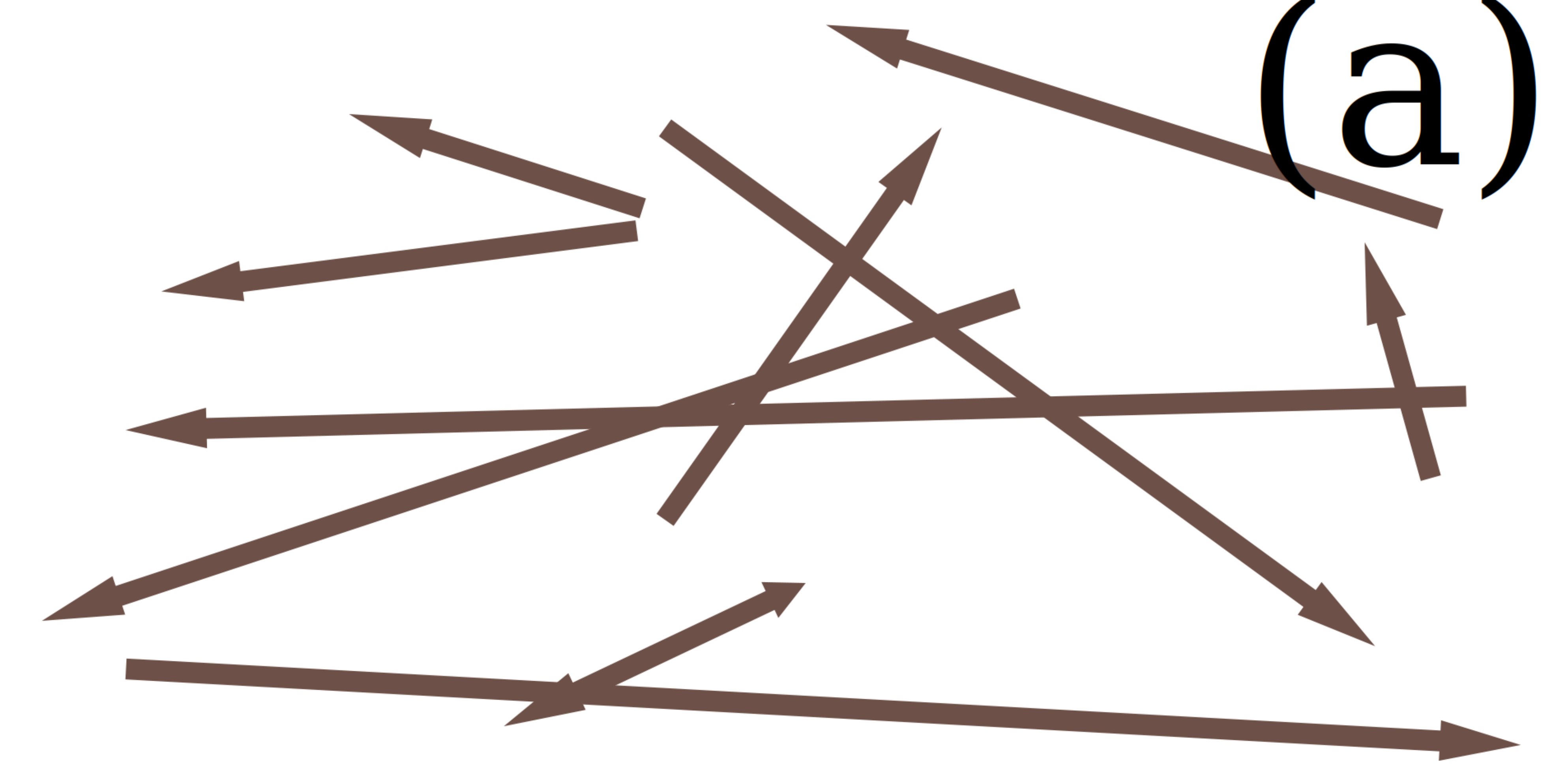}
    \includegraphics[width=.23\linewidth]{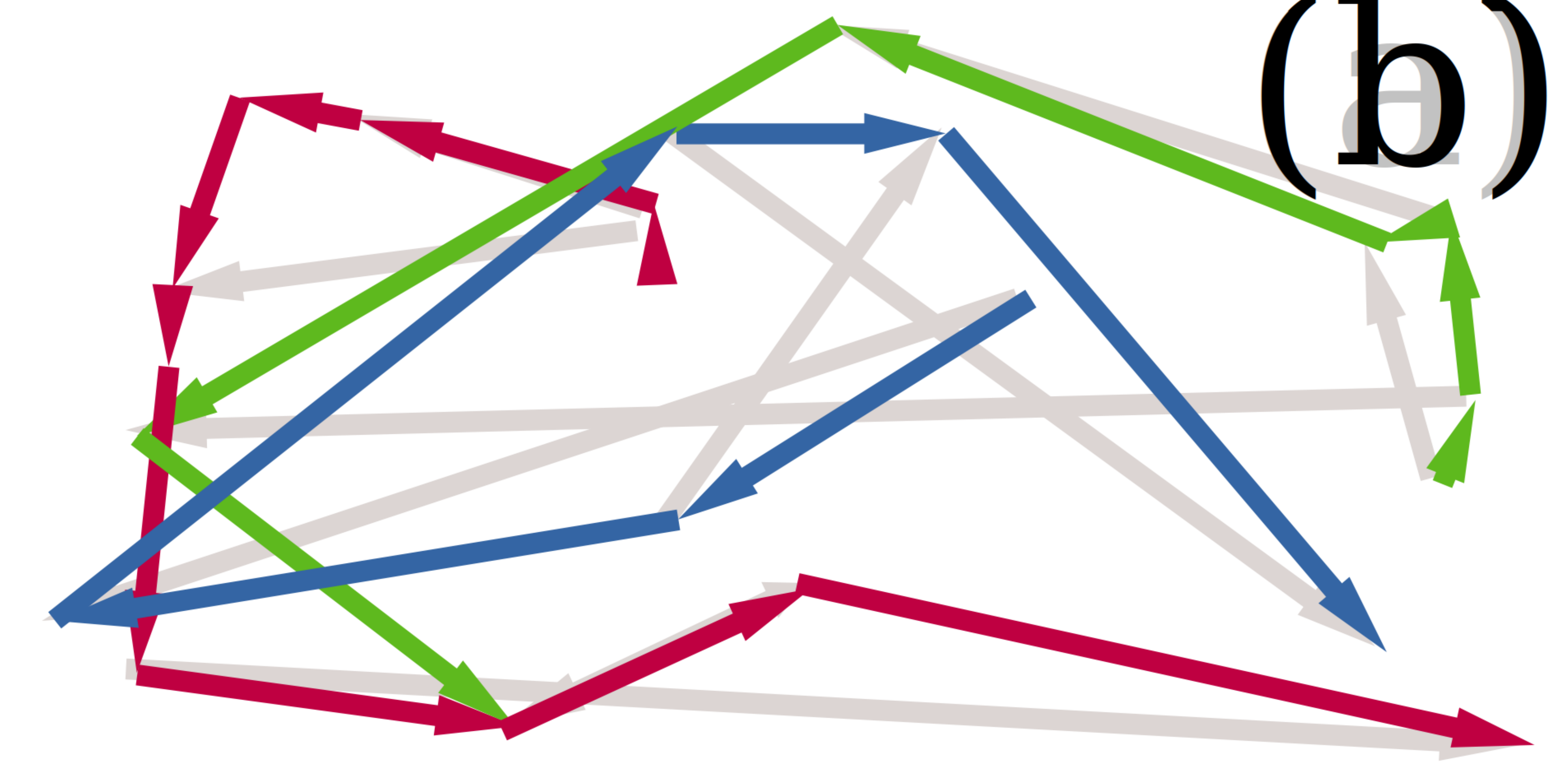}
    \includegraphics[width=.23\linewidth]{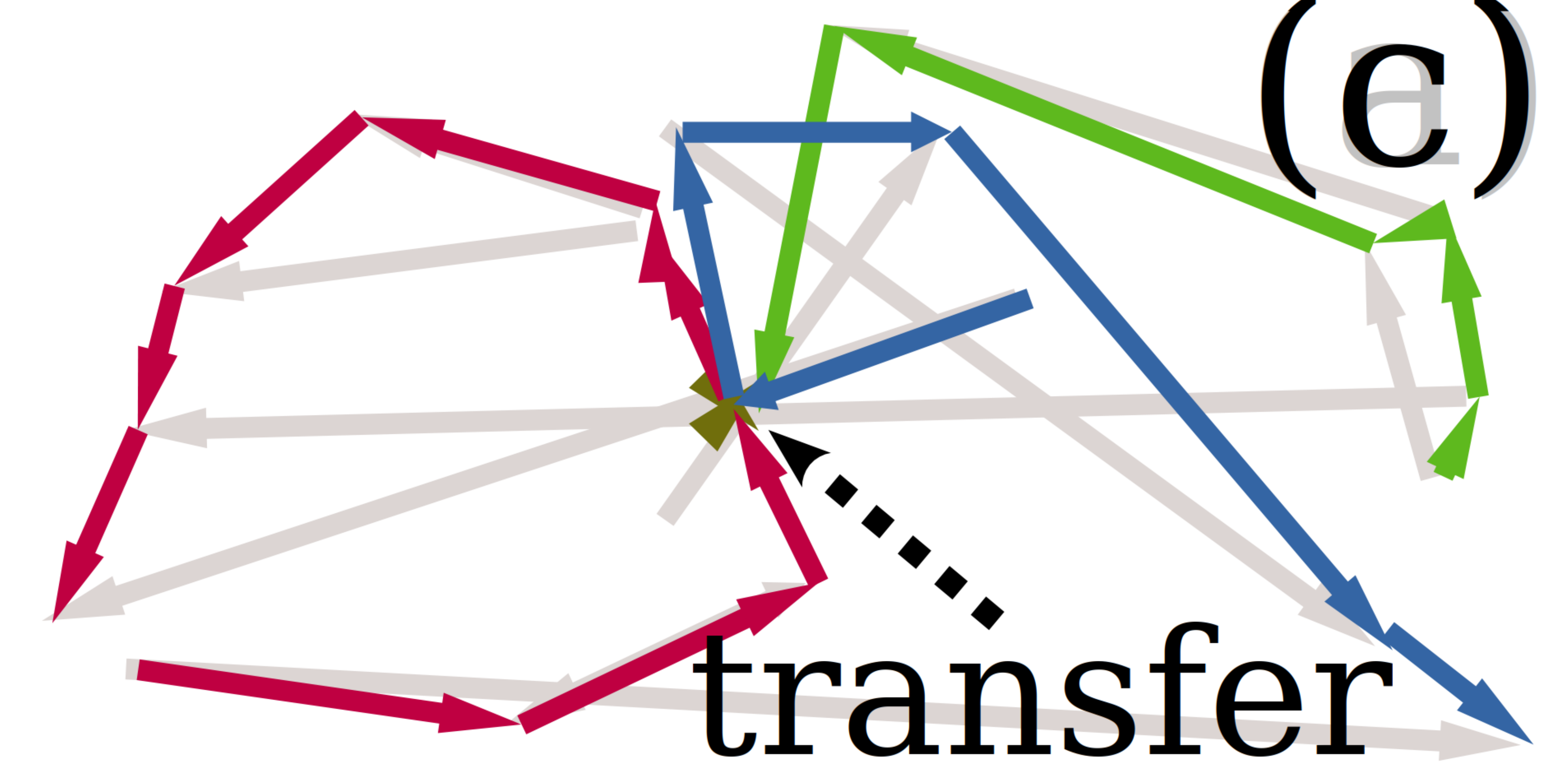}
    \caption{(a)~Trip requests (b)~DVRP-based routes (c)~Structured network of lines}
    \label{fig:example}
\end{figure}

Conventional public transport (CPT) systems rely on fixed lines and schedules, enabling strong demand consolidation and high efficiency. However, they are designed \emph{offline}, prior to operation, based on nominal conditions and demand. CPT thus lacks flexibility and offers poor service outside dense areas. In contrast, demand-responsive transport (DRT), typically managed via solving Dynamic Vehicle Routing Problems (DVRPs), lacks \emph{structure}: vehicle trajectories are adjusted individually, without an overall master plan. This often prevents the sharing of miles traveled among many users, leading to low occupancy and necessitating excessive subsidies, thereby hampering their sustainability \citep{currie2020most}. The importance of structure to achieve efficiency is illustrated in Fig.~\ref{fig:example}.

The most notable attempts to reconcile efficiency and adaptivity are now briefly discussed.
Hybrid systems combine CPT as a backbone and DVRP-based DRT in the last mile~\citep{calabro2023adaptive,fielbaum2024design}, but the respective limitations persist in both systems. 
\emph{Offline} systems improve efficiency by allowing transfers \cite{peton2014dial} or by
constructing bus lines based on user requests \cite{tong2017customized}. However, they require knowing the entire demand before computing routes, with prebookings the day before or longer. \citet{Ketter2023} design lines based on VRP, thereby preventing complex user journeys, which limits efficiency.

This paper proposes a novel way to operate DRT: instead of adjusting individual vehicle trajectories to incoming requests, we design a network of lines, which, as CPT, allows high demand consolidation and complex user journeys including transfers. Different from CPT, in which the network is designed \emph{offline} during strategic and tactical planning, we instead construct it \emph{online}, during its operation, dynamically adjusting it to incoming requests, which do not need to be known in advance.

%%%%%%%%%%%%%%%%%%%%%%%%%%%%%%%%%%%%%%%%%%
%%%%%%%%%% METHODOLOGY %%%%%%%%%%%%%%%%%%%
%%%%%%%%%%%%%%%%%%%%%%%%%%%%%%%%%%%%%%%%%%
\section{Methodology}

\subsection{Model}

We model the network of bus lines as an overlay graph~$\pazocal G(t)=(\pazocal E(t), \pazocal V), t\in[0,T]$. $\pazocal G(t)$ is embedded onto a substrate graph~$\pazocal G_\text{substr}=(\pazocal E_\text{substr},\pazocal V)$, where~$\pazocal V$ is the set of candidate stops and~$(u,v)\in\pazocal E_\text{substr}\subseteq\pazocal V\times\pazocal V$ indicates that a bus is allowed to go directly from stop~$u$ to stop~$v$ and we denote by~$w_{u,v}\ge 0$ the required time.\footnote{
This movement might require traversing several links on the road network.
}
Overlay graph~$\pazocal G(t)$ is a multi-layer graph, where each layer represents a line~$l\in\pazocal L$:
\begin{align}
\pazocal G_l(t)=\left(\pazocal E_l(t), \pazocal V\right)
&& \pazocal E(t)=\bigcup_{l\in\pazocal L} \pazocal E_l(t).
\end{align}

We model $\pazocal G(t)$ as a \emph{time-expanded graph}, which commonly represents public transport schedules \cite[Fig.~2]{FORTIN201618}. $\pazocal G(t)$ is composed of \emph{time-expanded arcs} $e=(\tau,v,v')\in\pazocal E_l(t)$, each indicating that a movement of a bus is scheduled to depart at instant $\tau>t$ to go from node~$v$ to~$v'$ (where it will arrive at~$t+w_{v,v'}$.\footnote{
In the following, we will not specify ``substrate'' or ``time-expanded'' for graphs and edges where the context is unambiguous.
}
Unlike static public transport, these lines are not predefined but are progressively constructed during operation, whence their dependence on~$t$. The decision policy to construct them will be detailed in the next subsection.

Requests arrive over time. In particular, request~$r=(\tau,v,v')$ arrives at instant~$\tau$ to go from node~$v$ to~$v'$. If a feasible path for~$r$ exists in~$\pazocal G(t), t=\tau$, the request is considered served. A feasible path for~$r$ may traverse multiple lines. Within each line, such a path is composed of a sequence of consecutive\footnote{
``Consecutive'' means that the arrival time and node of one edge correspond to the departure time and node of the next. The mathematical definition of feasible path is in our extended report \cite[(3)-(8)]{WangExtended}.
} time-expanded edges. Such a path may also include transfers from a line and another at certain nodes with some non-negative waiting time. A feasible path must be no longer than~$L$ time units ($L=30$ minutes in the numerical results). 
Note that multiple feasible paths might be available, and we let the user choose among them based on their preferences.\footnote{
For simplicity, when measuring travel times, we consider shortest paths in the numerical results. Note that the service of a user depends on the existence of at least one path and is independent of the user's route choice.
} 

\begin{figure}[t]
\centering

\begin{subfigure}{0.18\linewidth}
    \centering
    \includegraphics[width=\linewidth]{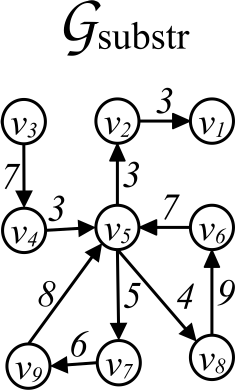}
    \label{fig:tega}
    (a)
\end{subfigure}
\hfill
\begin{subfigure}{0.47\linewidth}
    \centering
    \includegraphics[width=\linewidth]{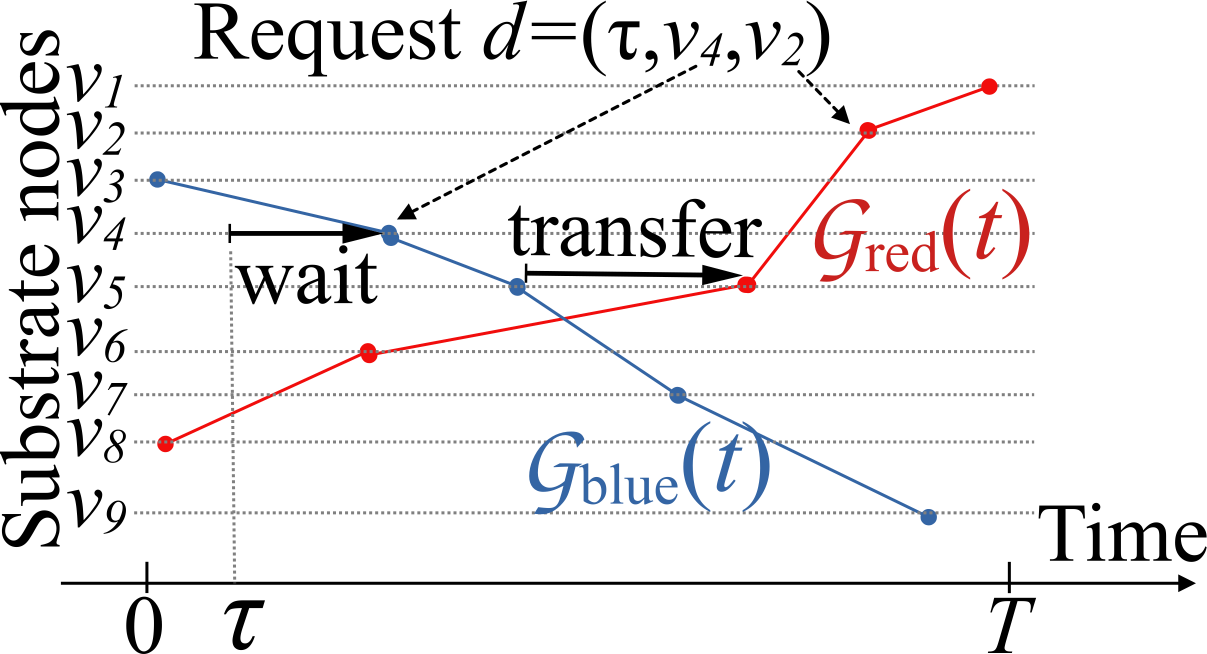}
    \label{fig:tegb}
    (b)
\end{subfigure}
\caption{On top of substrate graph~$\pazocal G_\text{substr}$ (a), we design time-expanded graph~$\pazocal G(t)$ composed of two layers (bus lines). Request~$\tau$ arrives at instant~$\tau$ and transfers from the blue to the red layer to get the service.}
\label{fig:sidebyside}
\end{figure}

This model is illustrated in Fig.~\ref{fig:sidebyside}.

If no feasible path exists in~$\pazocal G(t),t=\tau$, request~$r$ is placed in a buffer~$\pazocal B(t), t=\tau$. If subsequent evolution of~$\pazocal G(t), t>\tau$ yields a feasible path, $r$ is removed from~$\pazocal B(t)$ and marked as served.

The state of the system at time~$t$ is thus~$s(t):=\left( \pazocal G(t), \pazocal B(t) \right)$ and evolves due to requests arriving and being put in the buffer and to modifications of~$\pazocal G(t)$, as detailed in the subsection below.

\subsection{Network design decisions}
Designing the network of bus lines online consists of a sequence of decisions, each determining
\begin{itemize}
\item Which edge to add~$e=(\tau,v,v')$, i.e., which departure time~$\tau$, from which node~$v$ to which node~$v'$;
\item To which layer~$\pazocal G_l(t),l\in\pazocal L$ such an edge must be added, i.e., which bus must perform such movement;
\item At which instant~$t<\tau$ such an edge must be added. This is the time at which the corresponding bus movement is inserted in the plan (but the actual movement will occur at~$\tau>t$).
\end{itemize}

In our extended report \cite[Sec.~III.D]{WangExtended}, these decisions are formalized as an Impulsive Control Strategy of a Piecewise Deterministic Markov Process and we analytically prove that, if we aim to maximize the number of served requests, each edge~$e=(\tau,v,v')$ must be planned $L$ time units in advance, i.e., it must be added in overlay graph~$\pazocal G(t)$ exactly at time~$t=\tau-L$. We provide here an intuitive explanation. First, observe that for any movement departing at instant~$\tau$ from node~$v$ to~$v'$, it is preferable to plan it as soon as possible. Indeed, consider the corresponding time-expanded edge~$e=(\tau,v,v')$ and suppose it is added at time~$t$. The only requests that may exploit such an edge in their paths are those arriving in~$]t,\tau[$. Therefore, the sooner is the addition time~$t$, the higher is its contribution to serving requests, as this contribution spans a larger time interval. However, there is no reason to anticipate, earlier than~$\tau-L$, the insertion in the plan of a bus movement that will occur at instant~$\tau$. This is because we only consider as feasible the paths that are no longer than~$L$. Se the formal proof in \cite[Sec.~III.D]{WangExtended}.

We will thus only consider decision strategies where an edge~$e=(\tau,v,v')$ is inserted at instant~$t=\tau-L$. Establishing this exact timing, allows describing the decision strategy in terms of a Markov Decision Process over a discrete set of time steps~$\{t_n\}_{n=1,2,\dots}$, called intervention times, and with state~$s(t)=(\pazocal G(t_n), \pazocal B(t_n))$. At every intervention time~$t_n$, the action space and the next intervention time is found as follows:
\begin{enumerate}
\item For each line~$l\in\pazocal L$, we calculate its current termination time: suppose that the latest edge of a line~$l$ is $e=(\tau,v,v')\in\pazocal E_l(t)$, its arrival time~$\tau+w_{v,v'}$ corresponds to the termination time of that line.
\item We then find the ``shortest line'' (and denote with with~$l^*$), i.e., the one with the earliest termination time (which we denote with~$tt_l^*$). This is the line that needs to be extended before the others.
\item The action space is the set of potential time-expanded edges of layer~$\pazocal G_l(t)$ that depart at~$tt_l^*$. The edges of such action space are thus of type~$e_l^*=(tt_l^*, v_l^*,v_l^{\prime *})$, with~$(v_l^*,v_l^{\prime *})\in\pazocal E_\text{substr}$.
\item The next intervention time is set to~$t_{n+1}=tt_l^*-L$.
\end{enumerate}

The instantaneous reward is the number of requests served between~$t_n$ and~$t_{n+1}$, including both the requests that immediately find one or more feasible paths upon their arrival and the requests that were waiting in buffer~$\pazocal B(t)$ and that found a path later, thanks to the insertion of some later time-expanded edge. The objective is to maximize the cumulative reward, i.e., the total number of served requests.

Observe that, by construction, the four steps above ensure that at any instant~$t$, (i)~the current network~$\pazocal G(t)$ already contains the plan of all the movements up to instant~$t+L$, and (ii)~only modifications occurring after time~$t+L$ are allowed. This allows passengers arriving at~$t$ (i)~to plan their trips and (ii)~to have the guarantee that future network design decisions will not disrupt such trips.

\subsection{Solution method}
\begin{figure}
  \centering
  \includegraphics[width=0.35\linewidth]{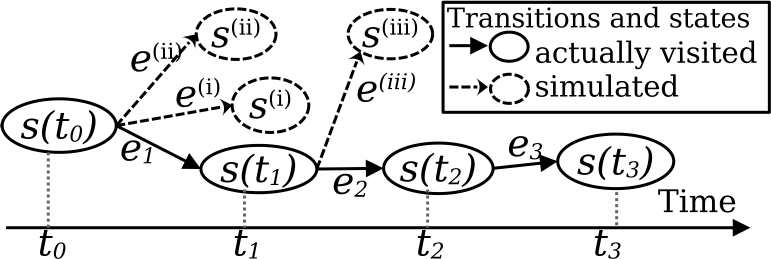}
  \caption{An example of search tree~$\textit{Tree}(s(t_0))$}
  \label{fig:Styles/NIPS_fig_1.png}
\end{figure}

The online network design decision problem described above has combinatorial complexity, due to the exploding number of possible sequences of edges to add. We use Monte Carlo Tree Search (MCTS) to approximate the optimal policy.
Each node in the search tree represents a partial network construction trajectory. From a given state, the algorithm explores possible sequences of actions by simulating future demand realizations (see depiction in Fig.~\ref{fig:Styles/NIPS_fig_1.png}). Note that during this exploration, the actual bus line network is not modified, and the sequences of actions are just simulated. During such a simulation, new future requests are simulated as well.
\footnote{
To keep our evaluation conservative, we avoided using advanced demand prediction models. In the numerical results, we instead adopted a simple predictor, consisting of a simple statistical model, calibrated on real taxi data, to capture the observed request arrival rate and geographical distribution of origins and destinations. This is detailed in our extended report~\cite[App.~III.A]{WangExtended}. Results are already good with this simple predictor, and even better results could be expected with more advanced forecast models.
}
In this way, we can account for the future rewards yielded by each action. 

The selection of actions (i.e., edges to be added) during tree expansion is guided by an exploration–exploitation trade-off. Let $Q(s,e)$ denote the estimated value of action $e$ in state $s$, and $N(s,e)$ the number of times it has been explored. Actions are selected according to:
\begin{equation}
e = \arg\max_{e} \left( Q(e,a) + c \cdot \pi^\text{aux}(e|s) \cdot \sqrt{\frac{\log N(s)}{N(s,e)}} \right),
\label{eq:MCTS}
\end{equation}
where $c$ is a parameter that prioritizes exploration of state-action pairs that have not been explored a lot.

The role of auxiliary policy~$\pi^\text{aux}(e|s)$ in the formula above is to further improve efficiency. Indeed, without it, we noticed that exploration was too dispersed, i.e., we were wasting simulation time trying unpromising network trajectories. To reduce this issue, we first ran our approach with a simplified version of~\eqref{eq:MCTS} (setting~$\pi^\text{aux}(e|s)=1,\forall e,s$). We dedicated many iterations (and thus much computation time) to explore possible network evolution trajectories. We then let the simplified version of our approach extract the promising trajectories. We collected hundreds of such exemplary trajectories in a training set. We then calibrated~$\pi^\text{aux}(e|s)$ as the likelihood of observing action~$e$ when in state~$s$ in the exemplary trajectories. After this preliminary calibration, we achieve that~$\pi^\text{aux}(e|s)$ gives higher weight to the best transitions and can be used in formula~\eqref{eq:MCTS} to bias exploration toward the promising state-action pairs. Therefore, the introduction of~$\pi^\text{aux}(e|s)$ constitutes an overhead that must be borne just once, for its calibration. After that, its use within~\eqref{eq:MCTS} is instantaneous. Full details of our solution method are provided in our extended version~\cite[Sec.~IV]{WangExtended}.

%%%%%%%%%%%%%%%%%%%%%%%%%%%%%%%%%%%%%%%%%%%%%%%%%%%%%%%
%%%%%%%%%%%%%%%%% RESULTS %%%%%%%%%%%%%%%%%%%%%%%%%%%%%
%%%%%%%%%%%%%%%%%%%%%%%%%%%%%%%%%%%%%%%%%%%%%%%%%%%%%%%
\section{Results and discussion}
\todoi{how comes that I ignore capacity? I remove requests waiting more than 30 minutes.}

\subsection{Considered scenario}
\label{sec:considered-scenario}

\noindent We evaluate our method against real user requests in Manhattan (Tab.~\ref{tab:param}). We replay a randomly selected 20\% subset of requests from public dataset~\cite{TLC_Trip_Record_Data}.
% Taxi zones are shown in Fig.~\ref{fig:taxi_zone_map_manhattan}. 
We use the centroids of the 67 taxi zones as candidate pickup/dropoff stops. 
% \todo{aa: I would indicate this the substrate graph with sth like~$\pazocal G_\text{substr}$, since we use~$\pazocal G_0$ already as root node of the tree. I would also use sth like~$\pazocal E_\text{substr}$ to indicate the edges of this graph.}
In substrate graph~$\pazocal{G}_\text{substr}$, set $\pazocal{V}$ contains such centroids, each of which is connected to all the others. Weight~$w_{u,v}$ of edge~$(u,v)\in\pazocal{E}_\text{substr}$ is the time for a bus to travel between nodes, i.e., $w_{u,v}=d(u,v)/v_\text{bus}$, where~$d_{u,v}$ is the distance between~$u,v\in\pazocal V$ and~$v_\text{bus}$ is the average speed of the bus.
The position of set~$\pazocal{L}$ of~$FS$ buses is randomly initialized.  
%We simulated our method from 9:00 to 13:00 on March 1, 2024. To get results faster, we used 20\% of the real request dataset from~\cite{TLC_Trip_Record_Data}.

% \todo{aa: when you use underscore, use compact notation. For instance, replace all $v_\text{car}$ with $v_\text{car}$ etc. Please do it for all these kinds of symbols}
%We train an auxiliary policy~$\pi_t^\text{aux}$ (\S{}\ref{sec:neural-policy}) on a training set constituted via applying vanilla MCTS on previous days of operation. We use it to improve our MCTS-based policy (\S{}\ref{sec:MCTS}), which we employ within Alg.~\ref{fig:algo_1}.

To train auxiliary policy~$\pi_t^\text{aux}$, we first construct a dataset as follows. We randomly sample 100 instances (100 days) of February 2024 from the dataset. For each sampled instance, we run a vanilla MCTS (with classical UCT scoring) between 9am: 1pm and record the resulting solution trajectory, each being a sequence of TEGs. We thus have 100 exemplary trajectories. At each decision step, we record the corresponding transition, i.e., which next stop of a certain bus is added to the trajectory. We also add into the input (i)~the ID of the stop where that bus is located before taking the action, (ii)~the travel times associated with all feasible actions (all the other stops), and (iii)~the cosine of the angle between each candidate action vector and the previous action of the same bus (we do this to let the policy learn to preserve a certain directionality of vehicle movements and avoid zig-zags). In this way, we extract movement patterns of good quality decisions from the exemplary trajectories, such as preferring next stops that are neither too close nor too far, and avoiding unnecessary detours and backtracking.

\begin{table}[t]
\centering
\begin{footnotesize}
  \begin{tabularx}{\columnwidth}{p{3.1cm}|X}
    \hline
    Parameter &Value\\
    \hline 
    Num. of candidate stops& 67\\
    Fleet size $FS$& 5, 10, 20, 30, 40, 50;
    Default (if not otherwise stated): 40
    \\
    Speed of a bus $v_\text{bus}$
    & 17.3 km/h \cite{speed_bus_taxis}
    \\
    Private car speed $v_\text{car}$
    & 11.4 km/h \cite{speed_bus_taxis}
    \\
    Walking speed $v_\text{walk}$
    & 4.3 km/h \cite{google_map1}
    \\
    Rollout termination& 5\\
    number $N_{end}$& \\
    Look-ahead buffer~$B$& 30 mins
    \\
    Training set to train~$\pi^\text{aux}$ 
    & All request data from~\cite{TLC_Trip_Record_Data}, February'24 
    \\
    Tested scenario 
    & 20\% of request data from \cite{TLC_Trip_Record_Data}, 9:00-13:00 
    March 1, 2024\\
   \hline
    
    \hline
  \end{tabularx}
\end{footnotesize}
  \caption{Scenario parameters}
  \label{tab:param}
\end{table}

\begin{table}[t]
\centering
\caption{Comparison with DVRP and CPT. Hyperparameters are in \emph{italic}.}
\label{tab:comparison}
\setlength{\tabcolsep}{2pt} % applies only within this table
\begin{minipage}{0.48\linewidth}
\centering
Dynamic Vehicle Routing Problem (DVRP)
\begin{tabular}{|l|c|c|c|c|}
\hline
 & \textit{Max wait} & Serv. & Trip & Wait \\
 &  \textit{time (min)} &  Rate & \multicolumn{2}{c|}{time (min)} \\
\hline
\multirow{2}{*}{SOTA} & \textit{20} & 34\% & 23 & 4 \\
                      & \textit{30} & 50\% & 25 & 5 \\
\hline
\multirow{2}{*}{\shortstack{Our\\method}} & \textit{20} & 63\% & 28 & 9 \\
                                          & \textit{30} & 90\% & 31 & 14 \\
\hline
\end{tabular}
\end{minipage}
\hfill
\begin{minipage}{0.48\linewidth}
\centering
Conventional Public Transp. (CPT)
\begin{tabular}{|c|c|c|c|c|c|}
\hline
 & \textit{Lines} & Bus/ & Serv. & Trip & Wait \\
 &       & line      &  rate  & \multicolumn{2}{c|}{time (min)} \\
\hline
\multirow{4}{*}{\rotatebox{90}{SOTA}} 
& \textit{4}  & 10 & 17\% & 10 & 1  \\
& \textit{10} & 4  & 50\% & 17 & 5  \\
& \textit{20} & 2  & 73\% & 32 & 16 \\
& \textit{40} & 1  & 80\% & 56 & 41 \\
\hline
\multicolumn{3}{|c|}{\shortstack{Our method}} & 91\% & 31 & 14 \\
\hline
\end{tabular}
\end{minipage}

\end{table}

\label{sec:results-bus}
% \todoi{aa: You should add another metric: the CDF of the spacing between a stop and another}
% \todoi{aa: You should add another metric: estimated operator cost spent per passenger.}
% \todo{aa: In Fig.~\ref{fig:3}-b, you should replace~``Nb of requests unserved'' with ``Fraction of requests unserved''}

We evaluate whether the proposed method yields a high \textbf{service rate} (more served requests) while using few resources (small \textbf{fleet size}), and providing a good user experience, by minimizing \textbf{trip time} (average \textbf{waiting time}+in-vehicle time). 
Since our service lays in the middle between dynamic taxis and and a conventional bus network, we compare it against 
\begin{itemize}
\item A State-Of-The-Art (SOTA) mobility on demand service management, based on the Dynamic Vehicle Routing Problem (DVRP), representing an optimized dynamic ride-haling service.
\item A SOTA conventional bus network, designed offline.
\item The performance of the real taxi service in Manhattan. 
\end{itemize}

\subsection{Comparison with State-Of-The-Art (SOTA) Dynamic Vehicle Routing Problem~(DVRP) solutions.}

In Table~\ref{tab:comparison}-left, we compare our solution with a well-known SOTA DVRP solution~\cite{alonso2017demand}, of which we run an open source implementation~\cite{Ride_Sharing}, with two different values of hyperparameter ``maximum waiting time'', fleet size of 40 vehicles for both solutions and infinite seat capacity.\footnote{We then verify that the actual vehicle occupancy achieved by our solution is compatible with buses and minibuses (Table~\ref{tab:3}).}

\begin{table*}[t]
    \centering
    \begin{tabular}{|c|c|c|c|c|}
        \hline                
                Fleet size & Avg. Occupancy 
                &  Avg. Stretch w.r.t & Avg. Stretch w.r.t &  Averrage number\\
                & (requests/bus)
                & direct car trip
                & walking trip
                & of transfers
                \\
         \hline
                  5 & 24&1.80& 0.68  & 1.24 \\ \hline
                  10& 14& 1.56& 0.61 &  1.79 \\ \hline
                  20& 9&  1.90&0.72  & 1.85  \\ \hline
                  30& 7&  2.35&0.89 &  1.59 \\ \hline
                  40& 6& 2.62&0.99 &  1.68 \\ \hline
                  50& 5& 2.83&1.07  & 1.50 \\ \hline
    \end{tabular}
    \caption{Performance with different fleet sizes }
    \label{tab:3}
\end{table*}

% We wish indeed to first evaluate the level of sharing the system can obtain and choose the appropriate seat capacity a-posteriori. 
Observe that the goal of SOTA~\cite{alonso2017demand} is to optimize an on-demand ``high-capacity'' ride-hailing. However, their DVRP-based approach limits the actual capacity they can sustain, whereas, via our network-centric approach, we nearly double the service rate, thanks to the efficiency brought by the network structure we design online. The conceptual reasons of such gains, explained in \S{}\ref{sec:introduction}, are confirmed by these results.\footnote{
Observe that both the SOTA DVRP solution and our method are applied on the same demand and substrate network~$\pazocal G_\text{substr}$. What changes between the two is the way vehicle movement is scheduled on top of~$\pazocal G_\text{substr}$: the former modifies vehicle routes at every request, while we design a ``masterplan'' time-expanded network~$\pazocal G(t)$ (overlay on top of~$\pazocal G_\text{substr}$), which vehicles will follow.
}
On the other hand, with our proposed system, users experience an increase in waiting and trip time and also need to deal with transfers. This is however likely acceptable, since it is similar to what they currently experience in Conventional Public Transport~(CPT).\footnote{
\label{fn:vs-taxis}
We do not aim to replace taxis, which would still be preferred by users willing to pay a high fee for maximum comfort; our system is rather a complement to CPT for everyday mobility. In other words, our method enables a brand-new transport service, rather than a ``better'' taxi service. Such a new service sits in the middle between the efficiency and high capacity of CPT and the adaptivity and comfort of taxis.
}

\subsection{Comparison with SOTA conventional bus network design} 

A key limitation of conventional public transport design is the need to fix the network structure a priori. In practice, this translates into selecting a set of lines, which determines the level of service coverage and operational cost.

The method of \cite{fielbaum2024design} represents conventional public transport (CPT), under the same setting (Table~\ref{tab:param}), and the same fleet size$=40$. Table~\ref{tab:comparison}-right shows that by increasing the number of lines, the static bus system covers more areas and improves the service rate. However, some areas remain consistently unreachable, resulting in 20\% unserved demand. Moreover, as the number of lines increases, trip and waiting times increase because fewer vehicles operate on each line. In contrast, our real-time design offers much higher service rates without excessively degrading trip and waiting times, due to its capability to adapt to the actual demand, rather than just operating lines previously planned offline.

\subsection{Comparison with real taxis}
From~\cite{TLC_Trip_Record_Data1} we measure that taxi real trips take in real-world $\sim 13$ minutes on average between 9am and 1pm, which are shorter than the ones our method provides (Table~\ref{tab:comparison}). However, we use only 40 buses, versus 2600 taxis and serve 90\% of the demand.\footnote{
Around 13k taxis operate in Manhattan~\cite{TLC_Trip_Record_Data1} and 2.6k correspond to the 20\% of 13k (remember that we considered 20\% of the real demand~Tab.~\ref{tab:param}).
} The loss of user-centric performance is thus justified by the much greater efficiency, which is compatible with the nature of the proposed system (see footnote~\ref{fn:vs-taxis}).

\subsection{Service characteristics}

\begin{figure}[]
  \centering
  \includegraphics[width=0.6\textwidth]{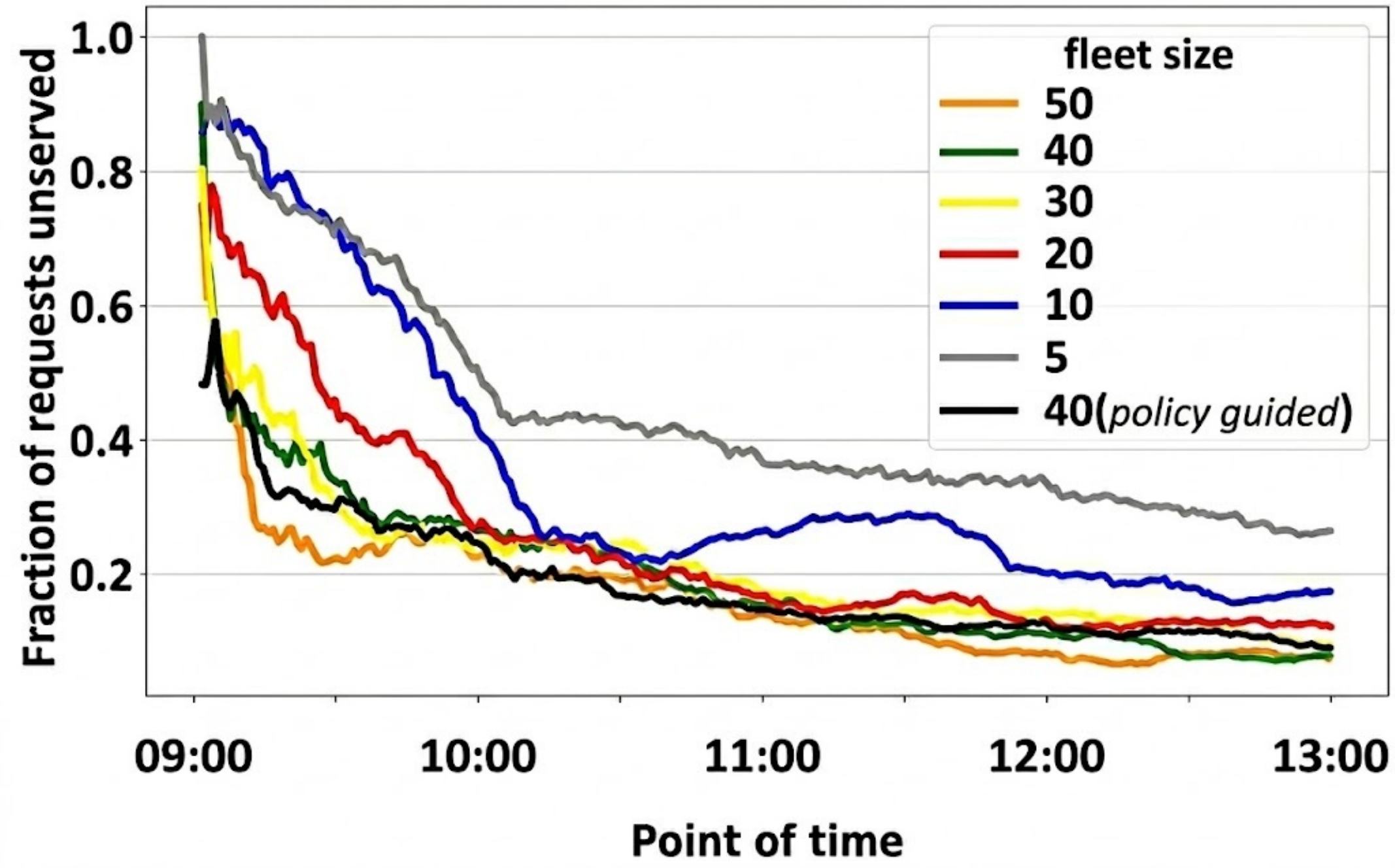}
  \caption{Fraction of requests unserved from 9:00 to 13:00.}
  \label{fig:3-1}
\end{figure}

\begin{figure}[]
  \centering
  \includegraphics[width=0.9\linewidth]{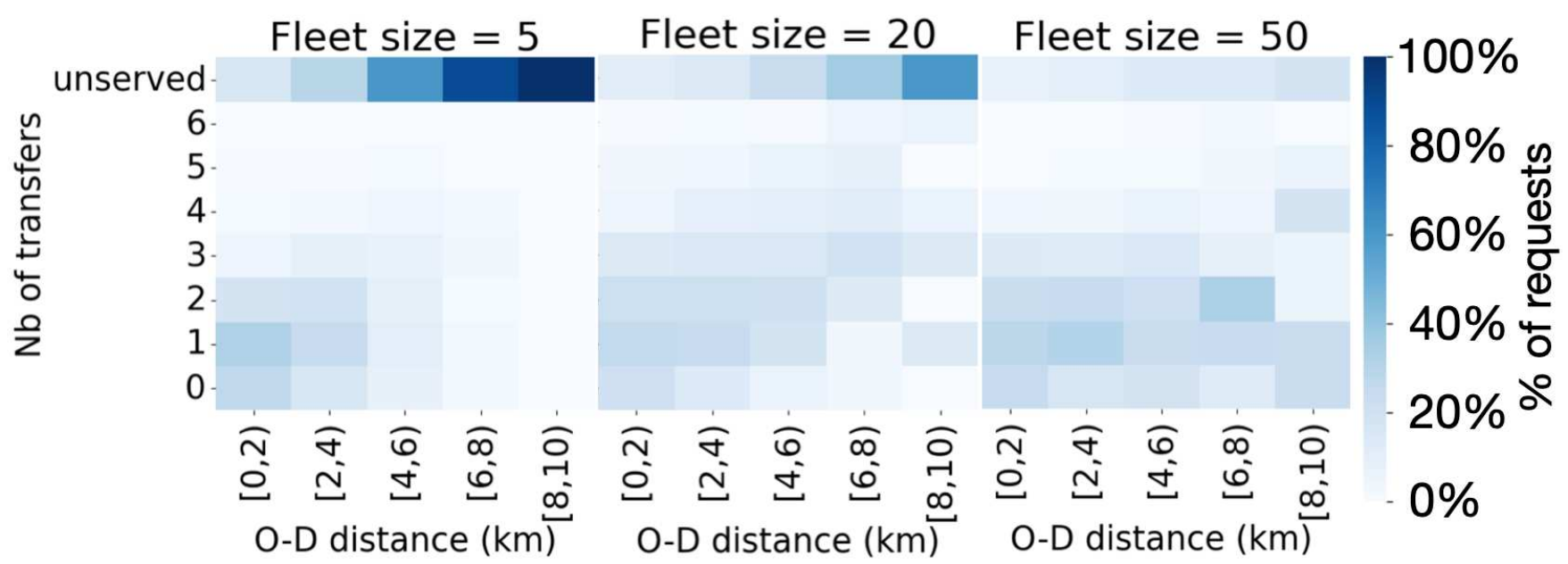}
  \caption{ Fraction of requests experiencing certain number of transfers over total requests, per each O-D distance interval. }
  \label{fig:heat_map_OD_distance_transfer}
\end{figure}

Fig.~\ref{fig:3-1} shows that the neural policy guidance is beneficial at the beginning, as it suggests directly good decisions, thus reducing from 40\% to 30\% unserved requests 30 minutes after startup, with fleet size~$\textit{FS}=40$. Fig.~\ref{fig:heat_map_OD_distance_transfer} shows that increasing fleet size allows serving longer trips, which mechanically results in a higher average trip time (Table~\ref{tab:2}).

\begin{figure}[h]
\centering
\includegraphics[width=0.9\linewidth]{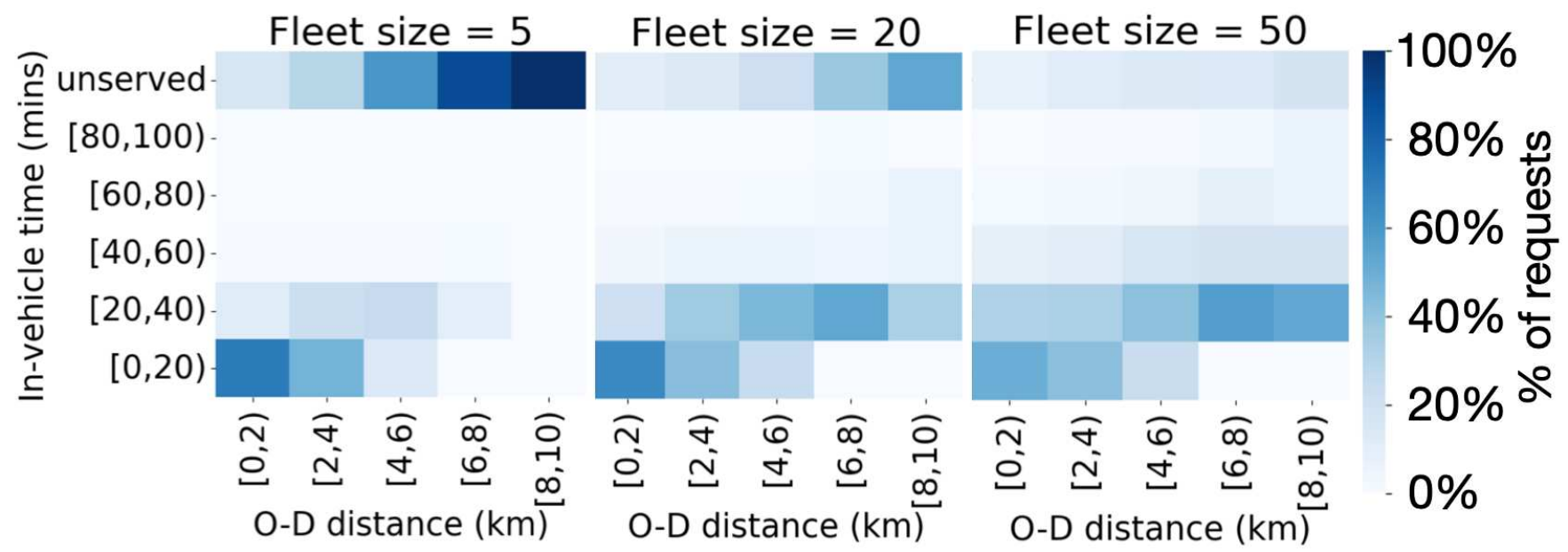}
\caption{Distribution of in-vehicle time by origin–destination distance}
\label{fig:in-vehicle}
\end{figure}

The in-vehicle time distribution (Fig.~\ref{fig:in-vehicle}) provides further insight into system performance. For short and medium distances, travel times remain close to those observed in taxi operations. For longer distances, travel times increase moderately due to transfers but remain significantly lower than those observed in static PT.

This result shows that the efficiency gains obtained through consolidation do not come at the expense of excessive detours. Instead, the system achieves a favorable balance between efficiency and user experience.

\begin{table*}[t]
    \centering
    \begin{tabular}{|c|c|c|c|c|}
        \hline                
                Fleet size 
                 &  Service rate &  Avg. Trip time (mins)&  Avg. Waiting time (mins)\\
         \hline
                  5 &  73.58\%&   27.81 &  16.01 \\ \hline
                  10 & 82.55\%&   27.75 &  15.76 \\ \hline
                  20 & 87.82\%&   29.68 &  15.03  \\ \hline
                  30 & 90.25\%&   29.93 &  14.20  \\ \hline
                  40 & 91.83\% &  31.79 &  14.63 \\ \hline
                  50 & 92.31\% &  32.20 &  13.89   \\ \hline
    \end{tabular}
    \caption{Comparison of results with different fleet sizes }
    \label{tab:2}
\end{table*}

Fig.~\ref{fig:heat_map_OD_distance_transfer} also shows that short trips are predominantly served without transfers. To adapt our approach to the case in which users accept no more than 2 transfers, we can label as infeasible any more complex path. In our current system, this would still preserve 88\% of completed requests.

\subsection{Real-time computation}
We pinpoint that our method meets the real-time requirements, since the choice of each action is made within seconds, i.e., much faster than the time between two consecutive intervention instants

\subsection{Discussion on the performance}

Table~\ref{tab:main} summarizes the main performance indicators of our method againts the SOTA. In summary, the proposed method achieves the best compromise between efficiency and service quality. It offers a high service rate with a small fleet size without excessively degrading trip and waiting times.

\begin{table}[h]
\centering
\begin{tabular}{lcccc}
\hline
Method & Fleet size & Service rate & Trip time & Wait time \\
\hline
DVRP (SOTA) & 40 & 50\% & 25 & 5 \\
Static PT (SOTA) & 40 & 80\% & 56 & 41 \\
Taxi (real Manhattan traces) & 2600 & 100\% & 18 & 2 \\
Proposed & 40 & 91\% & 31 & 14 \\
\hline
\end{tabular}
\caption{Comparison of global performance indicators}
\label{tab:main}
\end{table}

The results confirm that the performance gains stem from structural properties of the system.
First, DVRP is fundamentally limited by its local optimization structure, constrained by the trajectory-based formulation.
Second, static PT achieves consolidation but lacks adaptivity, which yields poor results under high demand variability.

The proposed approach combines both advantages by dynamically constructing a structured network. This enables demand consolidation while adapting to demand in real time.
Transfers play a central role in this mechanism. As shown in previous work, allowing transfers can significantly reduce detours and improve system efficiency.

Overall, the proposed paradigm resolves the classical trade-off between flexibility and efficiency in mobility systems.

\section{Conclusions}

This paper introduced a new paradigm for flexible mobility: the real-time design of public transport lines. By shifting from vehicle-level routing to network-level design, the proposed approach enables demand consolidation while retaining adaptivity.

Empirical results on real-world data show that the method significantly outperforms both DVRP-based approaches and static public transport design. These findings suggest that future mobility systems should move beyond purely demand-responsive routing and instead leverage adaptive network structures.

\bibliography{refs}

@article{FORTIN201618,
author={Fortin, Phillippe and Morency, Catherine and Trépanier, Martin},
title = {{Innovative GTFS Data Application for Transit Network Analysis Using a Graph-Oriented Method}},
journal = {J. Pub.Transp.},
year = {2016}
}

@misc{speed_bus_taxis,
  title = {New York City Mobility Report},
  author={Polly Trottenberg},
  year = 2018,
  howpublished = {\href{https://www.nyc.gov/html/dot/downloads/pdf/mobility-report-2018-screen-optimized.pdf}{[link]}}
}

@misc{TLC_Trip_Record_Data,
  title = {TLC Trip Record Data},
  author={NYC},
  year = 2024,
  howpublished = {\href{https://www.nyc.gov/site/tlc/about/tlc-trip-record-data.page}{[link]}},
}

@misc{TLC_Trip_Record_Data1,
  title = {Yellow Cab},
  author={NYC},
  year = 2024,
  howpublished = {\href{https://www.nyc.gov/site/tlc/businesses/yellow-cab.page}{[link]}},
}

@misc{google_map1,
  title = {Google Maps},
  author={Google},
  year = 2024,
  howpublished = {\href{https://www.google.com/maps}{[link]}}
}

@article{alonso2017demand,
  title={On-demand high-capacity ride-sharing via dynamic trip-vehicle assignment},
  author={Alonso-Mora, Javier and Samaranayake, Samitha and Wallar, Alex and Frazzoli, Emilio and Rus, Daniela},
  journal={Proc. Nat. Acad. Sci.},
  year={2017}
}

@misc{Ride_Sharing,
  title = {Ride Sharing},
  author={MetaZuo},
  year={2017},
  howpublished = {\href{https://github.com/MetaZuo/RideSharing}{[link]}}
}

@article{fielbaum2024design,
  title={Design of mixed fixed-flexible bus public transport networks by tracking the paths of on-demand vehicles},
  author={Fielbaum, Andres and Alonso-Mora, Javier},
  journal={Transp. Res. Part C},
  year={2024},
  publisher={Elsevier}
}

@misc{WangExtended,
  title        = {Online Design of Dynamic Networks - Extended version},
  author       = {Wang, Duo and Araldo, Andrea and El Yacoubi, Mounim},
  year         = {2026},
  eprint       = {2410.08875},
  archivePrefix= {arXiv},
  primaryClass = {cs.AI},
  url          = {https://arxiv.org/abs/2410.08875}
}

@article{currie2020most,
  title={Why most DRT/Micro-Transits fail--What the survivors tell us about progress},
  author={Currie, Graham and Fournier, Nicholas},
  journal={Res. Transp. Econ.},
  year={2020},
  publisher={Elsevier}
}

@article{calabro2023adaptive,
  title={Adaptive transit design: Optimizing fixed and demand responsive multi-modal transportation via continuous approximation},
  author={Calabr{\`o}, Giovanni and Araldo, Andrea and Oh, Simon and Seshadri, Ravi and Inturri, Giuseppe and Ben-Akiva, Moshe},
  journal={Transportation Research Part A: Policy and Practice},
  volume={171},
  year={2023}
}

@article{tong2017customized,
  title={Customized bus service design for jointly optimizing passenger-to-vehicle assignment and vehicle routing},
  author={Tong, Lu Carol and Zhou, Leishan and Liu, Jiangtao and Zhou, Xuesong},
  journal={Transportation Research Part C: Emerging Technologies},
  volume={85},
  pages={451--475},
  year={2017},
  publisher={Elsevier}
}

@article{peton2014dial,
  title={The dial-a-ride problem with transfers},
  author={Masson, Renaud and Lehu{\'e}d{\'e}, Fabien and P{\'e}ton, Olivier},
  journal={Computers \& Operations Research},
  volume={41},
  pages={12--23},
  year={2014},
  publisher={Elsevier}
}

@article{Ketter2023,
  title={Balancing convenience and sustainability in public transport through dynamic transit bus networks},
  author={Abdelwahed, Ayman and van den Berg, Pieter L and Brandt, Tobias and Ketter, Wolfgang},
  journal={Transportation Research Part C: Emerging Technologies},
  volume={151},
  pages={104100},
  year={2023},
  publisher={Elsevier}
}

\end{document}